\documentclass[a4paper,12pt]{article}
\usepackage[pctex32]{graphics}

\begin{document}
\topmargin 0pt \oddsidemargin 0mm

\renewcommand{\thefootnote}{\fnsymbol{footnote}}
\begin{titlepage}
\begin{flushright}
INJE-TP-05-02\\
hep-th/0501118
\end{flushright}

\vspace{5mm}
\begin{center}
{\Large \bf Holographic energy density in the Brans-Dicke theory}
\vspace{12mm}

{\large  Hungsoo Kim, H. W. Lee and Y. S. Myung\footnote{e-mail
 address: ysmyung@physics.inje.ac.kr}}
 \\
\vspace{10mm} {\em  Relativity Research Center and School of
Computer Aided Science, Inje University Gimhae 621-749, Korea}
\end{center}

\vspace{5mm} \centerline{{\bf{Abstract}}}
 \vspace{5mm}
We study  cosmological applications of the holographic energy
density. Considering the holographic energy density as a dynamical
cosmological constant, we need the Brans-Dicke theory as a
dynamical framework instead of general relativity. In this case we
use the Bianchi identity as a consistency relation to obtain
physical solutions. It is shown that the future event horizon as
the IR cutoff provides the dark energy in the Brans-Dicke theory.
Furthermore the role of the Brans-Dicke scalar is clarified  in
the dark energy-dominated universe by calculating its equation of
state.
\end{titlepage}
\newpage
\renewcommand{\thefootnote}{\arabic{footnote}}
\setcounter{footnote}{0} \setcounter{page}{2}
\section{Introduction}
Supernova (SN Ia) observations suggest that our universe is
accelerating and the dark energy contributes $\Omega_{\rm
DE}\simeq 0.60-0.70$ to the critical density of the present
universe~\cite{SN}. Also  cosmic microwave background (CMB)
observations~\cite{Wmap} imply that the standard cosmology is
given by the inflation and FRW universe~\cite{Inf}. A typical
candidate for the dark energy is the cosmological constant in
general relativity. Recently Cohen {\it et al} showed that in
quantum field theory, the UV cutoff $\Lambda$ is related to the IR
cutoff $L_{\rm \Lambda}$ due to the limit set by forming a black
hole~\cite{CKN}. In other words, if $\rho_{\rm \Lambda}$ is the
quantum zero-point energy density caused by the UV cutoff, the
total energy of the system with size $L_{\rm \Lambda}$ should not
exceed the mass of the system-size black hole: $L_{\rm \Lambda}^3
\rho_{\rm \Lambda}\le L_{\rm \Lambda}/G$. Here the Newtonian
constant  $G$ is related to the Planck mass ($G=1/M_p^2$). The
largest $L_{\rm \Lambda}$ is chosen as the one saturating this
inequality and its holographic energy density is then given by
$\rho_{\rm \Lambda}= 3c^2M_p^2/8\pi L_{\rm \Lambda}^2$ with a
factor $3c^2$. Taking $L_{\rm \Lambda}$ as the size of the present
universe (Hubble horizon: $R_{\rm HH}$), the resulting energy  is
comparable to the present dark energy~\cite{HMT}. Even though this
holographic approach leads to the data, this description is
incomplete because it fails to explain the equation of state for
the dark energy-dominated  universe~\cite{HSU}. In order to
resolve this situation, one  introduces another candidates for the
IR cutoff. One is the particle horizon $R_{\rm PH}$. This provides
$\rho_{\rm \Lambda} \sim a^{-2(1+1/c)}$ which gives the equation
of state $\tilde{\omega}_{\rm \Lambda}=1/3$ for $c=1$~\cite{LI}.
It corresponds to a radiation-dominated universe and unfortunately
it is a decelerating universe. In order to find an accelerating
universe, we need the future event horizon $R_{\rm FH}$. In the
case of $L_{\rm \Lambda}=R_{\rm FH}$ one finds $\rho_{\rm \Lambda}
\sim a^{-2(1-1/c)}$ which describes the dark energy with
$\tilde{\omega}_{\rm \Lambda}=-1$ for $c=1$. This is close  to the
data~\cite{SN}. The related works appeared in
ref.\cite{FEH,Myung2,Med}.

However, the above choices for the IR cutoff have something to be
needed to clarify. $\rho_{\rm \Lambda}= 3M_p^2/8\pi L_{\rm
\Lambda}^2$ with $L_{\rm \Lambda}=R_{\rm HH},R_{\rm PH},R_{\rm
FH}$ correspond to dynamical cosmological constants. But authors
investigated its cosmological applications in the framework of
general relativity. We need the dynamical framework, for example
the Brans-Dicke theory, to study the cosmological application of
the dynamical cosmological constant.

In this work we ask how  the dark energy can be realized from the
Brans-Dicke theory with the holographic energy density. In the
framework of general relativity, the choice of IR cutoff as the
Hubble horizon did not give us a correct equation of state.
Furthermore a recent work on this direction did not use the
Bianchi identity as a consistency relation to derive the dark
energy from the Brans-Dicke theory with holographic energy
density~\cite{Gong}. If one does not use the Bianchi identity, it
is not easy to discriminate between physical and unphysical
solution.  We show that the future event horizon as the IR cutoff
provides the dark energy in the Brans-Dicke theory with the
holographic energy density. Here we use all of equations including
the Bianchi identity as a consistency relation~\cite{KimSS}. These
are consisted of four equations but it reduces to three equations
after eliminating the pressure by making use of the
energy-momentum conservation  for a matter-fluid. In addition, the
role of the Brans-Dicke scalar is clarified by deriving its
equation of state.

\section{Brans-Dicke theory with a perfect fluid}
We start with the action for the Brans-Dicke theory with a perfect
fluid\cite{KimSS}
\begin{equation}
\label{action} {\cal S}_{BD}=\int
d^4x\sqrt{-g}\Big[\frac{1}{16\pi} \Big(\phi R-\omega
\frac{\nabla_{\mu}\phi \nabla^{\mu}\phi}{\phi} \Big)-{\cal
L}_m(\psi,g_{\mu\nu})\Big] \end{equation} in the Jordan frame.
Here $\omega$ is a parameter to be determined.  Their equations
are given by
\begin{eqnarray}
\label{EQTN-1}&& G_{\mu\nu}=8\pi
T_{\mu\nu}^{BD}+\frac{8\pi}{\phi}T_{\mu\nu}^{m},\\
\label{EQTN-2}&& \nabla^2\phi=\frac{8\pi}{2\omega+3}T^{m~
\lambda}_{\lambda},
\end{eqnarray}
where the energy-momentum tensor $T^m_{\mu\nu}$ for a matter-fluid
and $T^{BD}_{\mu\nu}$ for a Brans-Dicke scalar take the forms
\begin{eqnarray}
\label{emtm-1}
&& T^m_{\mu\nu}=(\rho+p)u_{\mu}u_{\nu}+pg_{\mu\nu}, \\
\label{emtm-2} && T^{BD}_{\mu\nu}=
\frac{1}{8\pi}\Big[\frac{\omega}{\phi^2}\Big(\nabla_{\mu}\phi\nabla_{\nu}
\phi-\frac{1}{2}g_{\mu\nu}(\nabla
\phi)^2\Big)+\frac{1}{\phi}\Big(\nabla_{\mu}\nabla_{\nu}\phi-g_{\mu\nu}
\nabla_{\alpha}\nabla^{\alpha}\phi\Big)\Big],\\
\label{emtm-3}&&\tilde{T}^{BD}_{\mu\nu}=
(\rho_{BD}+p_{BD})u_{\mu}u_{\nu}+p_{BD}g_{\mu\nu} \equiv
T^{BD}_{\mu\nu}/G_0.
\end{eqnarray}
Here the last expression corresponds to the definitions  of energy
density and pressure for the Brans-Dicke scalar with a constant
$G_0$.
 Let us introduce  a
$(3+1)$-dimensional Friedman-Robertson-Walker (FRW) metric
\begin{equation}
\label{2eq1} ds^2 =-dt^2 +a(t)^2 \Big[ \frac{dr^2}{1-kr^2} +r^2
d\Omega^2_{2} \Big],
\end{equation}
where $a$ is the  scale factor of the universe and $d\Omega^2_{2}$
denotes the line element of a two-dimensional unit sphere. Here
$k=-1,~0,~1$ represent that the universe  is open, flat, closed,
respectively. Assuming that $\phi=\phi(t)$, a cosmological
evolution is determined by the four equations\cite{Gong,KimSS}
\begin{eqnarray}
\label{2eq2-1} && H^2 +H
\frac{\dot{\phi}}{\phi}-\frac{\omega}{6}\Big(\frac{\dot{\phi}}{\phi}\Big)^2
 =\frac{8\pi}{3 \phi}\rho
-\frac{k}{a^2},\\
\label{2eq2-2}&& \ddot{\phi}+3H\dot{\phi} =-\frac{8\pi}{2\omega+3}(\rho -3p),\\
\label{2eq2-3}&& \dot{\rho}+3H(\rho+p)=0,\\
\label{2eq2-4}&&\dot{\rho}_{BD}+3H\Big(\rho_{BD}+p_{BD}\Big)=
\frac{\rho}{G_0}\Big(\frac{\dot{\phi}}{\phi^2}\Big).
\end{eqnarray}
Here $H={\dot a}/a$ represents the Hubble parameter and the
overdot stands for  derivative with respect to the cosmic time
$t$. The first equation corresponds to  the Friedmann equation,
the second comes from the equation (\ref{EQTN-2}) for a
Brans-Dicke scalar, and the third is the conservation law
$\nabla_{\nu}T^{m \mu \nu}=0$ for a matter-fluid. The last comes
from the Bianchi identity of $\nabla_{\nu}G^{\mu\nu}=0$ and plays
a role of the consistency relation~\cite{KimSS}. If a solution
does not satisfy the  four equations
Eqs.(\ref{2eq2-1})-(\ref{2eq2-4}) simultaneously, it is no longer
a physical solution. In order to investigate a role of the
Brans-Dicke scalar, one needs  its energy density and pressure
derived from Eqs.(\ref{emtm-2}) and (\ref{emtm-3})as
\begin{eqnarray}
\label{2eq4}
 && \rho_{BD}=\frac{1}{16 \pi G_0}\Big[\omega
 \Big(\frac{\dot{\phi}}{\phi}\Big)^2-6H\Big(\frac{\dot{\phi}}{\phi}\Big)\Big],\\
&& p_{BD}=\frac{1}{16 \pi G_0}\Big[\omega
\Big(\frac{\dot{\phi}}{\phi}\Big)^2+4H\Big(\frac{\dot{\phi}}{\phi}\Big)+2\Big(\frac{\ddot{\phi}}{\phi}\Big)\Big].
\end{eqnarray}
In the Brans-Dicke scalar with $\rho=0$, from Eq.(\ref{2eq2-4})
its equation of state is given by $p_{\rm BD}=\tilde{\omega}_{\rm
BD}\rho_{\rm BD}$ with $\tilde{\omega}_{\rm BD}=-\frac{1}{3}$.
This means that a Brans-Dicke scalar by itself gives zero
acceleration. Thus one expects that it plays an intermediate role
between matter and cosmological constant.

\section{Brans-Dicke theory with Holographic energy density}

The Brans-Dicke scalar $\phi$ plays a role the inverse of
Newtonian constant ($\phi \sim 1/G$). In this case we have a
relation of $\phi_0 \sim 1/G_0$. For definiteness we choose the
holographic dark energy with $c=1$ as\cite{LI,Gong}
\begin{equation} \label{hde}\rho_{\rm \Lambda}=\frac{3\phi}{8\pi
L^2_{\rm \Lambda}}.
\end{equation}
From now on we explore the role of the Brans-Dicke scalar in the
holographic energy-dominated universe with $k=0$. First we choose
the IR cutoff as the present universe-size (Hubble horizon: HH)
such that $L_{\rm \Lambda}=1/H$. Then one finds the power-law
solutions from Eqs.(\ref{2eq2-1})-(\ref{2eq2-4}) as
\begin{equation}
a(t)=a_0t^{\omega/(4\omega+6)},~~ \phi(t)=\phi_0t^{3/(2\omega+3)},
~\rho_{\rm \Lambda}=\frac{3\omega^3 \phi_0}{8\pi
(4\omega+6)^2}\Big(\frac{a}{a_0}\Big)^{-2(4\omega+3)/\omega}.
\end{equation}
We note that  the Bianch identity in Eq.(\ref{2eq2-4}) is
satisfied with any $\omega$~\cite{KimSS}. The last relation
implies the equation of state defined by $p_{\rm
\Lambda}=\tilde{\omega}^{\rm HH}_{\rm \Lambda} \rho_{\rm
\Lambda}$,
\begin{equation}
\tilde{\omega}^{\rm HH}_{\rm \Lambda}=-1
+\frac{2(4\omega+3)}{3\omega}.
\end{equation}
On the other hand the Brans-Dicke scalar gives
\begin{equation}
\rho^{\rm HH}_{\rm BD}=0,~p^{\rm HH}_{\rm BD}=\frac{6}{8\pi G_0}
\frac{H^2}{\omega}
\end{equation}
which provides a zero energy density and positive pressure.   In
the general relativistic limit of $\omega \to \infty$, one finds
$\rho_{\rm BD}=p_{\rm BD}=0$ which means that the Brans-Dicke
scalar does not play any role.

On the other hand, one recovers the equation of state in the limit
of $\omega \to \infty$ as
\begin{equation} \tilde{\omega}^{\rm HH}_{\rm \Lambda} \to
\frac{5}{3}.
\end{equation}
This equation of state is regarded as the correct one for the
Hubble horizon. We remark the general relativistic equation of
state. The Friedmann equation $H^2=8\pi G\rho_{\rm \Lambda}/3$
with $\rho_{\rm \Lambda}=3c^2H^2/8\pi G$ does not give the
equation of state. This gives nothing but $c=1$. Further,
combining the matter density $\rho_{\rm m}=a_0/a^3$ with
holographic energy density $\rho_{\rm \Lambda}$, its Friedmann
equation takes the form of  $\rho_{\rm m}=3(1-c^2)H^2/8\pi G$.
This implies that $\rho_{\rm m}$ behaves as $H^2~(\rho_{\rm
\Lambda})$~\cite{HSU,LI}. It  leads to a dust-like equation of
state: $\tilde{\omega}^{\rm HH}_{\rm \Lambda}=0$. Especially for
$c=1$, one cannot find any information. This means that the
equation of state for the holographic energy density with $L_{\rm
\Lambda}=R_{\rm HH}$ cannot be determined within the framework of
general relativity. In this case the Brans-Dicke theory is more
favorable to fixing the equation of state. Actually it is given
not by $\tilde{\omega}^{\rm HH}_{\rm \Lambda}=0$ but an
decelerating universe with $\tilde{\omega}^{\rm HH}_{\rm
\Lambda}=5/3$. However, we want to find the accelerating universe
with $\omega<-1/3$ and thus this
 is not the case.

 In order to
resolve this situation, one is forced to introduce the particle
horizon (PH): $L_{\rm \Lambda}=R_{\rm PH}=a \int_0^t (dt/a)=a
\int^a_0(da/Ha^2)$.  We assume the power-law solutions of
$a(t)=a_0t^r,~\phi=\phi_0t^s$ with $\rho_{\rm
\Lambda}=\frac{3\phi}{8\pi R^2_{\rm PH}}$. In this case, an
accelerating solution is possible for $r>1$. One obtains three
relevant equations from Eqs.(\ref{2eq2-1}), (\ref{2eq2-2}), and
(\ref{2eq2-4}) after eliminating the pressure $p$ using
Eq.(\ref{2eq2-3}) as\cite{Gong}
\begin{eqnarray}
\label{parho-1}
&& (s+2)r=\frac{\omega s^2}{6}+1,\\
\label{parho-2}&& (2\omega+3)(3r+s-1)rs=(r-1)^2(12r+3s-6),\\
\label{parho-3}&& rs(\omega+1)-2r^2+3r-\frac{\omega s}{3}=1,
\end{eqnarray}
where the last relation corresponds to the Bianchi
identity~\cite{KimSS}. From $\rho_{\rm \Lambda}$ the equation of
state is given by
\begin{equation}
\tilde{\omega}^{\rm PH}_{\rm \Lambda}=-1-\frac{s-2}{3r}.
\end{equation}
 Their solutions which satisfy Eqs.(\ref{parho-1}) and (\ref{parho-2}) are given by
\begin{eqnarray}
\label{solPH}
 && (1)~ r=\frac{1}{2},~s=0,~\omega={\rm arbitrary},~\tilde{\omega}^{\rm PH}_{\rm \Lambda}=1/3 \nonumber \\
&& (2)~ r=-\frac{s}{4}+\frac{1}{2},~s={\rm
arbitrary},~\omega=-\frac{3}{2},~\tilde{\omega}^{\rm PH}_{\rm
\Lambda}=1/3 \nonumber
\\
&& (3)~ r={\rm
arbitrary},~s=-3+\frac{1}{r},~\omega=-\frac{6r^3}{(3r-1)^2},~\tilde{\omega}^{\rm
PH}_{\rm \Lambda}=-1+\frac{5r-1}{3r^2} \nonumber
\\&& (4)~ r={\rm
arbitrary},~s=-2r,~\omega=- \frac{3(2r^2-2r+1)}{2r^2}. \nonumber
\end{eqnarray}
The first three cases satisfy also the Bianchi identity
Eq.(\ref{parho-3}) as a consistency relation. (1) case recovers
the general relativistic solution with the equation of state
$\tilde{\omega}_{\rm GR}=1/3$ in the limit of  $\omega \to
\infty$. (2) corresponds another radiation-dominated solution in
the Brans-Dicke theory with the holographic energy density. In the
case of (3), one finds the dark energy equation of state
$\tilde{\omega}_{\rm \Lambda} \to -1$ as $r\to \pm \infty$. The
case of $r<0$ is compatible  with  $\omega\to \infty$ which
correspond to the general relativistic case.  However, considering
$a(t)=a_0 t^{r}$ with $r<0$ leads to a contracting universe. Thus
this case is not an interesting solution. In the case of
$\omega<0~(r>0)$, one finds a negative Brans-Dicke term  which
does not lead to  general relativity in the limit of $r \to
\infty$.  However, we find  an accelerating phase with
$\tilde{\omega}^{\rm PH}_{\rm \Lambda}< -1/3$ for  $r \ge 3$. On
the other hand the BD energy density and pressure are given by
\begin{equation}
\label{BRPH}\rho^{\rm PH}_{\rm BD}=-\frac{3(2r-1)}{8 \pi G_0
r^2}H^2,~P^{\rm PH}_{\rm BD}=\frac{1}{8\pi G_0
r^4}\Big(-9r^3+14r^2-7r+1\Big)H^2.
\end{equation}
 The solution  (4) satisfies Eqs.(\ref{parho-1}) and
(\ref{parho-2}) only but it does not satisfy the Bianchi identity
Eq.(\ref{parho-2}). Hence it is excluded for our purpose.

Finally one  introduces the future event horizon (FH): $L_{\rm
\Lambda}=R_{\rm FH}=a \int_t^{\infty} (dt/a)=a
\int_a^{\infty}(da/Ha^2)$. This was used in the holographic
description of cosmology by Li~\cite{LI}. Here we assume the
power-law solutions of
$\phi/\phi_0=(a/a_0)^{\alpha},~H/H_0=(a/a_0)^{\beta-1}$ with
$\rho_{\rm \Lambda}=\frac{3\phi}{8\pi R^2_{\rm FH}}$. Here we
require $\beta \ge 1$ to find a physical solution. In this case,
one obtains three relevant equations from Eqs.(\ref{2eq2-1}),
(\ref{2eq2-2}), and (\ref{2eq2-4}) after eliminating the pressure
$p$ using Eq.(\ref{2eq2-3})\cite{Gong}
 \begin{eqnarray}
\label{feh-1}
&& \beta^2=1+\alpha-\frac{\omega}{6}\alpha^2,  \\
\label{feh-2}&&(2\omega+3)(\alpha+\beta+2)\alpha=3\beta^2(\alpha+2\beta+2),\\
 \label{feh-3}&& \Big(\frac{\omega
 \alpha}{3}-1\Big)\beta+\frac{\alpha}{3}(2\omega+3)=\beta^2,
\end{eqnarray}
where the last equation comes from the Bianchi
identity\cite{KimSS}. The equation of state is given by
\begin{equation}
\tilde{\omega}^{\rm FH}_{\rm
\Lambda}=-1-\frac{\alpha+2\beta-2}{3}.
\end{equation}
 Their solutions which satisfy Eqs.(\ref{feh-1}) and (\ref{feh-2}) are given by
\begin{eqnarray}
 && (1)~ \alpha=-2(\beta+1),~\beta={\rm arbitrary},~\omega=-3/2,~\tilde{\omega}^{\rm FH}_{\rm \Lambda}=1/3 \nonumber \\
&& (2)~ \alpha=0,~\beta=-1,~\omega={\rm
arbitrary},~\tilde{\omega}^{\rm FH}_{\rm \Lambda}=1/3 \nonumber\\
&& (3)~ \alpha=(\beta+2)(\beta-1),~\beta={\rm
arbitrary},~\omega=\frac{6}{(\beta+2)^2(\beta-1)}, \nonumber \\
 &&~\tilde{\omega}^{\rm FH}_{\rm \Lambda}=-1-\frac{(\beta-1)(\beta+4)}{3}
 \nonumber\\
 && (4)~ \alpha=-2,~\beta={\rm arbitrary},~
 \omega=-\frac{3}{2}\Big(\beta^2+1\Big). \nonumber
\end{eqnarray}
The first three cases satisfy also the Bianchi identity
Eq.(\ref{feh-3}) as a consistency relation. The solution (1) is a
radiation-dominated phase for the Brans-Dicke theory with the
holographic energy density. (2) corresponds another
radiation-dominated solution in the Brans-Dicke theory with the
holographic energy density. In the case of (3), one finds the dark
energy equation of state $\tilde{\omega}^{\rm FH}_{\rm \Lambda}
\to -1$ when $\omega\to \infty~(\beta\to 1)$. Actually this
corresponds to the dark energy solution of the general
relativistic case with the future event horizon. On the other
hand, the Brans-Dicke energy density and pressure are given by
\begin{equation}
\label{BDFH} \rho^{\rm FH}_{\rm BD}=-\frac{3(\beta^2-1)}{8 \pi
G_0}H^2 \le 0,~P^{\rm FH}_{\rm BD}=\frac{(\beta-1)}{8\pi
G_0}\Big(\beta^3+4\beta^2+3\beta+1\Big)H^2\ge 0, \end{equation}
where the equalities hold when $\omega\to \infty~(\beta\to 1)$. We
emphasize that the Bianch identity Eq.(\ref{2eq2-4}) as a
consistency relation  plays a important role in  the Brans-Dicke
theory combined with the holographic energy density. However, (4)
case does not satisfy the Bianchi identity and thus it is not the
case.

\section{Discussions}

We discuss the role of  the holographic energy density with the IR
cutoff in the Brans-Dicke theory. It is very interesting to
investigate the role of  dynamical cosmological constant
(holographic energy density) in the dynamical framework
(Brans-Dicke theory). This may  be considered  as an analogy that
the role of cosmological constant is usually investigated in the
framework of general relativity. In this case, one has to use
three equations from Eqs.(\ref{2eq2-1}), (\ref{2eq2-2}), and
(\ref{2eq2-4}) after eliminating the pressure $p$ using
Eq.(\ref{2eq2-3}). Here Eq.(\ref{2eq2-4}) is the Bianchi identity
as a consistency relation.

In the case of the Hubble horizon as the IR cutoff, its equation
of state is not fixed in  general relativity because of  handicap
of the framework\cite{HSU,LI}. However, one finds that
$\tilde{\omega}^{\rm HH}_{\rm \Lambda}\to 5/3$ in the $\omega \to
\infty $ limit of the Brans-Dicke theory. It represents the
correct equation of state for the holographic energy density with
the Hubble horizon as the IR cutoff. However, it derives a
decelerating universe.

We find that the particle horizon as the IR cutoff could provide
two radiation-dominated phases as well as an accelerating phase in
the Brans-Dicke theory. But we cannot recover the dark energy with
$\tilde{\omega}^{\rm PH}_{\rm \Lambda} \to -1$ in the general
relativistic limit. Hence the particle horizon is not suitable for
a candidate of the IR cutoff to obtain the dark energy.

In the case of  the future event horizon as the IR cutoff we find
a phantom-like equation of state $\tilde{\omega}^{\rm FH}_{\rm
\Lambda}\le -1$ from the Brans-Dicke theory with holographic
energy density. In the limit of $\omega \to \infty$, we
immediately find the equation of state  $\tilde{\omega}^{\rm
FH}_{\rm \Lambda} \to -1$ for a  general relativistic case.

Finally we comment on the role of the Brans-Dicke scalar in view
of the holographic energy density. In the case of the  future
event horizon, we have $\rho_{\rm \Lambda}>0,~p_{\rm \Lambda}<0$
while  $\rho_{\rm BD} \le 0,~p_{\rm BD} \ge 0$.  This means that
the Brans-Dicke scalar plays a different role when comparing to
that of the holographic energy density. Actually the Brans-Dicke
scalar slows down the expansion-rate of the holographic energy
density as the dynamical cosmological constant~\cite{KimSS}. This
is because $p_{\rm tot}=p_{\rm \Lambda}+p_{\rm BD}$ becomes less
negative. As the time goes on, the role of the Brans-Dicke scalar
is negligible, while the holographic energy density derive an
accelerating universe solely.

\section*{Acknowledgment}
H.W. was in part supported by KOSEF, Astrophysical Research Center
for the Structure and Evolution of the Cosmos.

\end{document}